# Flash Crash for Cash:
# Cyber Threats in Decentralized Finance

Kris Oosthoek
*Cyber Threat Intelligence Lab*
*Delft University of Technology*
Delft, The Netherlands
k.oosthoek@tudelft.nl

*Abstract*—Decentralized Finance (DeFi) took shape in 2020. An unprecedented amount of over 14 billion USD moved into DeFi projects offering trading, loans and insurance. But its growth has also drawn the attention of malicious actors. Many projects were exploited as quickly as they launched and millions of USD were lost. While many developers understand integer overflows and reentrancy attacks, security threats to the DeFi ecosystem are more complex and still poorly understood. In this paper we provide the first overview of in-the-wild DeFi security incidents. We observe that many of these exploits are market attacks, weaponizing weakly implemented business logic in one protocol with credit provided by another to inflate appropriations. Rather than misusing individual protocols, attackers increasingly use DeFi's strength of permissionless composability against itself. By providing the first holistic analysis of real-world security incidents within the nascent financial ecosystem DeFi is, we hope to inform threat modeling in decentralized cryptoeconomic initiatives in the years ahead.

*Index Terms*—cryptoeconomics, decentralized finance, smart contracts, cyber security, threat intelligence

## I. INTRODUCTION

More than ten years after the introduction of Bitcoin as a peer-to-peer digital cash system, its societal impact is reflected in the pioneering of central banks with Central Bank Digital Currency, as well the International Monetary Fund's announcement of a new Bretton Woods for 2021. Institutional players have recognized the staying power of 'programmable' digital money and explore its application to their benefit. But other than currency, blockchains also hold potential for other financial applications with which the open-source cryptoeconomic community has moved forward.

Decentralized Finance (DeFi) aims to disentangle and provide freedom of choice in financial services. It aspires to replace central bureaucratic institutions with computer code, stored in a smart contract and available on the blockchain for inspection by everyone. As smart contracts are application-agnostic and permissionless, DeFi allows everyone with development competencies to interact with and build financial systems. Clients don't have to trust the promises or reputation of a middleman, only the protocol. Redeeming control from the middleman provides freedom, but comes with a responsibility. A user can only hold himself accountable for costly mistakes. '*Don't trust, verify*' in crypto parlance. Each transaction must be verified as users can't outsource trust to a middleman.

The 'DeFi summer' of 2020 reflected the increased excitement of investors for DeFi's potential, as the total value locked in DeFi projects grew from 1 billion to 12 billion USD in a few months time. But many DeFi projects were attacked as quickly as they launched. Much of the dynamic was reminiscent of the 2017 ICO hype, showing that many DeFi projects are high-risk ventures from a financial and a technological perspective. Due to a fear of missing out accelerated by media coverage of high yields, many unskilled investors lost money. While speculative investing is implicitly financially risky, risk in these projects was also non-financial. Millions of USD value were put into new, unaudited smart contracts of developers with Twitter handles stating they 'test in prod' [1]. Many projects were launched with a focus on speed and agility and less on security. Audit processes were omitted and funds were lost due to a lack of basic security hygiene. Incidents also occurred with projects audited by multiple reputable security companies, which simply failed to observe critical issues.

Awareness of 'technical' software weaknesses in Ethereum's smart contract language Solidity, such as reentrancy, access control and integer overflow vulnerabilities is now widespread due to early attacks on The DAO, Parity Wallet and ERC-20 tokens respectively. While still a significant issue, this area has seen significant academic contributions [2]–[5] and has improved much. Technical vulnerabilities in smart contracts are relatively easily mitigated if adhering to community-audited ERC-20 and ERC-721 contract standards instead of coding from scratch. In DeFi, it appears that attackers prefer other avenues of exploitation, misusing legitimate Ethereum and DeFi functionality. Projects are primarily hacked through misuse of the economic functionality that they seek to disrupt. Decentralized capabilities such as voting, arbitrage and flash loans are exploited and have cascading effects through interdependencies between DeFi protocols, known as 'money lego'. A DeFi derivatives platform might depend on third-party services to supply the bid-ask prices, but a vulnerability allowing to read price data during a transaction may introduce significant financial risk.

As opposed to traditional finance, developers of DeFi projects often lack financial experience and cyber security is an afterthought in a hasty development process. The hosting of smart contract code on Ethereum further enables an attacker's opportunity to locate vulnerabilities quickly and efficiently. Exploiting DeFi projects currently is a low-risk high reward opportunity to malicious actors. Where Ethereum has seen isolated smart contract attacks in the past, over the 'DeFi summer' of 2020 adversaries 'flash crashed' compositions of DeFi protocols and cashed out lavishly. As the increase in private and institutional investor interest in cryptoeconomics will also attract malicious actors, it is key to extract cyber threat intelligence from these events. By exploring in-the-wild security incidents with DeFi in this paper and proposing a classification scheme, we make the following contributions:

- We give an overview of reported attacks on DeFi projects to explore how malicious actors are exploiting DeFi.
- We show that the major attacks in terms of financial impact employed compositions of DeFi protocols.
- We provide a framework of cyber security threats to DeFi projects to standardize the discussion on threats to DeFi.
- We provide an attack life cycle of flash-loan funded attacks, the ubiquitous attack vector in 2020.
- Per threat in the framework we provide guidance on how these can be mitigated in protocol development.

## II. AN OVERVIEW OF DECENTRALIZED FINANCE

Decentralized Finance (DeFi), sometimes called Open Finance, is the cryptoeconomic system experimenting with alternative financial services. It aims to do disrupt finance by replacing the security and integrity provided by traditional central bookkeepers such as banks with trustless, transparent protocols executed by a smart contract. For many financial services such as exchanges, lending, derivatives, payments and assets, a decentralized offering already exists. The DeFi community is also experimenting with more niche applications such as prediction markets and margin trading.

Currently most value is locked within DeFi protocols focused at lending and borrowing, in which users deposit an asset such as ETH to take a loan in another asset, such as a stablecoin. Other protocols such as Uniswap and Curve run the decentralized exchange (DEX) of assets, as an alternative to centralized counterparts such as Coinbase. Yield farming, prompted by Yearn, can be compared to highly active asset management as traditionally performed by hedge funds. Furthermore price oracles, which provide 'ground truth' price information to DeFi protocols, are not so much end-user protocols but critical DeFi infrastructure.

In 2020 the total value locked in DeFi projects has reached an all-time high of 14.744 billion USD on December 1, 2020 [6]. The largest financial market in the world, currency trading on the foreign exchange market accounted for $2,409,000,000 ($2.409 quadrillion) in the same year. Proportionally as well as nominally DeFi is still insignificant. Although 'DeFi' is likely obsolete terminology soon, its fundamental dynamics play a key role in the disruption of global financial services.

DeFi initiatives are built on the Ethereum blockchain and use smart contracts to implement financial protocols. The encoding of financial protocols in smart contracts provides trustlessness, permissionless and interoperable provision of eventual financial transactions. Transactions are not managed by a central institution but transparently deployed on a public blockchain. There is no middleman to be trusted, only a protocol (*code is law*). It also allows everyone to create financial systems of their own or to participate in those offered by others. New applications can be built by composing multiple DeFi protocols, also known as 'money lego'. The DeFi space is largely experimental and still in its nascency. It is forgotten quickly that projects aren't final, stable financial products yet.

The yield farming hype in 2020, is exemplary of DeFi's current state. Many of experimental projects resulted in high gains to early investors, which thus attracted speculative investors. The high initial yields of few projects led to copycat projects being launched with food names like Sushi, Yam, Spaghetti and Kimchi. Basically attempts to 'get rich quick' for both aspiring developers, but also speculative investors who dumped money into a project without a white paper or website. As many of these projects were forks of other immature and untested projects with minor tweaks, this introduced major attack surface. While many of these initiatives experienced a short live span, it is also normal behavior of more risk-oriented actors early in the technology adoption life cycle.

While DeFi aims to provide the seamless user experience associated with ATMs and store POS as currently provided by traditional players, currently its user experience is still relatively spartan. But apart from accessibility challenges due to its experimental nature, DeFi has systematic challenges it needs to solve. Apart from the financial risk of investing in DeFi, there are many facets to non-financial risk. Examples are inadequate storage of private admin keys, causing single points of failure and thus centrality. Many DeFi projects also inherit risk from the projects on which they are building (composability), or run the risk of receiving prices from vulnerable or simply dishonest price oracles. Furthermore, risk is caused by developers launching DeFi protocols while lacking financial literacy, let alone the risk of protocols getting shut down through regulatory action. DeFi is a space with many threats, both from a financial and non-financial perspective. As however usual with early innovations, security is not a primary concern. Where names such as Dridex, ZeuS, Carbanak and the Lazarus Group have become household names for actor groups targeting traditional financial institutions, this intelligence is absent in DeFi. Therefore it is crucial to retrieve a full holistic overview of in-the-wild security threats to DeFi.

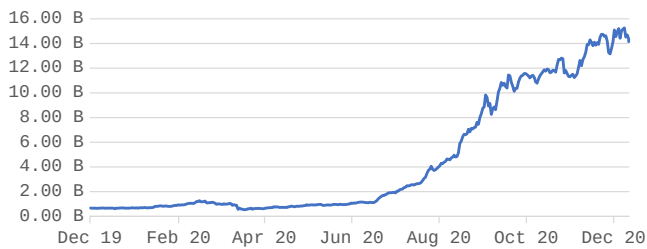

Fig. 1. Total Value Locked in DeFi Protocols December 2019-2020

## III. Related Work

Extensive research has been performed in the area of smart contracts. From previous work a lot is known about vulnerabilities and weaknesses in smart contract code, while research on DeFi attacks specifically is still in its early stages.

### A. Smart Contract Attacks

Research into smart contract security threats is interchangeably referred to as smart contract security and smart contract risk [7], [8]. An early overview of security incidents with smart contracts was provided by Buterin [9]. Praitheeshan et al. have focused on vulnerabilities inherent to smart contracts on Ethereum and surveyed older incidents, such as the DAO and Parity multisig attacks [5]. Chen et al. have performed a similar survey and provide a classification of Ethereum vulnerabilities and appropriate defenses [10]. Several authors have focused on historical attacks on the DAO, Governmental, King of the Ether Throne and mitigations against those [2], [4], [11], [12]. Other authors have surveyed specific vulnerabilities such as reentrancy [13], [14] and integer overflows [15], [16]. Similar to the Common Weakness Enumeration (CWE) for application software, initiatives have been launched for a category system of Solidity weaknesses, such as SWC Registry [17] and the Decentralized Application Security Project [18]. Dingman et al. have classified smart contract weaknesses using NIST's Bugs Framework [19]. These however capture Solidity vulnerabilities, not the recent attack vectors observed in DeFi.

An adjacent field focuses on security analysis through static testing and formal analysis. Harz and Knottenbelt have surveyed high-level smart contract languages and their security characteristics, as well verification methods [3]. Other authors have analyzed the security of smart contracts using symbolic execution and static analysis tools [20]–[23]. Such mechanisms inherently leave some real-world attack vectors unaddressed. Perez and Livshits however regarded vulnerabilities against their real-world impact and found that of 23,327 vulnerable contracts, only 1.98% had been exploited [24].

### B. DeFi Protocol Attacks

Related work on our primary research topic, exploitation of DeFi protocols, is relatively limited due to the short existence of the industry. Significant work has been performed by Daian et al., who studied arbitrage bots in DEXes 'front running' transactions of ordinary users through priority gas auctions to manipulate order transaction execution [25]. Qin et al. were the first to explore the phenomenon of flash loans [26], which first surfaced in 2020 and is also present in our analysis. Gudgeon et al. have published a governance exploit in the Maker protocol [27]. By exploiting the its governance system, they were able to increase Maker's token supply. From an economic security perspective, Gudgeon et al. found that liquidity in three of the current major lending platforms in DeFi, as few as three accounts controlled the majority of the total liquidity [28]. Furthermore Klages-Mundt have regarded economic and security risks to custodial and non-custodial stablecoins [7].

## IV. Methodology

Our research goal was to provide insight into the nature of security threats to DeFi projects. We did this by building a dataset of vulnerabilities exploited by malicious actors as well verified attack vectors disclosed by security researchers. Based on observations from this dataset, we have developed a framework to understand how these threats relate to other cryptoeconomic and Ethereum threats.

*Data Collection:* A dataset of incidents was gathered in November 2020 using Google Custom Search API queries *defi attack* and *defi hack*. In addition *defi* abbreviated, these queries were also executed for *decentralized finance*. The interim dataset was cross-checked against the *href* attributes of each result to discover cross-links to incidents missed initially. Automated search queries on popular crypto websites *cointelegraph.com* and *coindesk.com* were used as a final check for completeness. Using word frequency analysis, we found an initial 26 incidents.

Only incidents with protocols listed on DeFi Pulse [6] were included in our sample. DeFi Pulse is the de facto standard source for DeFi market information and is referenced in related academic contributions [27], [28]. Listing requires a whitepaper, Github repo, Twitter profile and project page thus filters out scams, which excluded 5 projects.

Furthermore we have only included incidents with on-chain contracts. Through the inclusion of on-chain contracts, we exclude vulnerabilities found in code hosted in repositories, which are not a direct threat to a project's continuity. This requirement excluded the double spend bug in SushiSwap's governance, discovered in code hosted in the official Github repository. With this requirement, the eventual amount of incidents in our dataset was 20. Addresses of affected contracts are available for reference in Table I.

*Attack Classification:* As we found the terminology on attacks DeFi to be fuzzy and non-standardized, we have developed a framework to structure the discussion on security threats to DeFi projects and show how these differ from attacks to individual smart contracts. The framework is informed by previous work in the area of smart contract security, as well security incidents in DeFi recorded in our dataset. We have plotted the attacks in our dataset on this framework based on information available by post-mortems from a primary source, as well transaction actions on Etherscan.

*Financial Impact:* We have used Etherscan and official post-mortems to establish the financial impact of security incidents.

## V. Attacks on DeFi Projects

In this section we enumerate attack techniques used in DeFi security incidents, which include exploits by malicious actors or attack vectors disclosed by security researchers.

Table I provides an overview of the attacks discussed in chronological order. From left to right, the *Date* column provides the date of the incident. The *Project* column lists the name of the affected project, the *Type* column the financial service it provides. The *Contract* column lists the address of the contract running the affected DeFi protocol. The *Attack* and

*Technique* columns list attack type and technique. *USD Impact* shows the financial impact of incidents in which funds were stolen. Funds retrieved from or returned by the attacker are subtracted in order to accurately reflect total value lost. The *Source* column references first-hand incident reports, which were used to classify the attack and provide background on the tools, techniques and procedures (TTPs) employed.

### A. Market Attacks

The attack vector in 10 of the total 20 incidents in our sample targeted DeFi market mechanisms by exploiting compositions of multiple DeFi protocols. In accordance with observations by other authors [25], [26] we find that flash-loan funded price oracle attacks are a significant attack vector. Out of the 10 attacks that exploited compositions of multiple protocols, all but one were flash loan-funded price oracle attacks. The prevalence and novelty of this specific category of attacks is further discussed in the next section.

*1) Price Oracle Exploits:* In total 9 projects in our sample were attacked through their implementation of a price oracle. Of these, 6 projects were exploited by malicious actors and consequently lost funds. As price oracles are usually exploited with flash loans, both concepts are considered briefly below.

**Price Oracle Exploits:** in order to exploit a vulnerable implementation of a price data feed into a DeFi protocol, adversaries need to fabricate arbitrage opportunities. By swapping large amounts of tokens, usually on a DEX or with a liquidity pool, they create price variations to create an arbitrage opportunity. As many DeFi projects depend on a 'price oracle' - a price data feed from another project to establish market price. Especially when depending on a feed from a single source, the dependency this creates can be catastrophic to a project. Flash loans are a welcome vehicle to facilitate large loans that trigger substantial liquidity changes.

**Flash Loans:** a DeFi concept without an equivalent in traditional finance, these are loans with a term of a single blockchain transaction. They allow borrowers to pursue arbitrage opportunities in the time span of the transaction. Lenders, usually assembled in a liquidity pool, run no risk of borrowers defaulting, as the transaction will fail if the lender does not pay back. Hence flash loans are non-collateralized as they do not require advance collateral from the debtor. First introduced by Aave, flash loans provide an opportunity for the creation of novel financial products [48].

**Flash Loan-Funded Price Oracle Exploits:** flash loans provide a vehicle to 'weaponize' attacks to amplify profit. The only limitation is that all the attack steps should be executed within a single block. In case of these attacks, the flash loan is however just part of a bigger attack vector. It also requires vulnerable smart contract code or business logic, the flash loan is used to multiply the potential outcomes of exploitation.

In February 2020, bZx, a margin trading and lending service, was attacked twice on two consecutive days using a flash loan-funded price oracle attack. The attackers used the flash loans to manipulate wBTC and sUSD prices. As bZx relied on a single price oracle exclusively [33], [34], the price information from this single source was what finalized the attack. A similar attack vector was disclosed to Nexus Mutual, which uses a price oracle to trigger a re-balance via Uniswap of the holdings of its mutual fund. The callback function to the oracle however was implemented such that everyone could trigger the re-balance and interact with Uniswap on behalf of this function [36]. Balancer suffered an attack in which an attacker took a flash loan of wrapped ether, which then was repeatedly swapped for a deflationary token [41]. In October, Harvest was attacked used a flash loan to funnel 24 million USD from liquidity pools used as vaults by the protocol. This was the biggest attack in terms of attacker yield to date, due to an 'engineering error' that allowed to bypass a checking function. The smart contract was independently audited by three reputable auditors [45]. Half November, Akropolis was attacked with a flash loan-funded attack that exploited an unchecked token whitelist used for price oracle input handling, to drain its 2 million Dai holdings [46]. Value was attacked through a vulnerability in its deposit handling [47].

Both DDEX and bZx were vulnerable to price manipulation as they used price oracles without validating the price returned. Both attack vectors were disclosed by a security researcher in September 2019 [31], [33]. While not exploited, both could have been weaponized using flash loans.

Flash loans provide low risk high reward opportunities to attackers, as they allow them to attack without upfront cost. If the attack fails, the flash loan simply reverts at the end of the block time with only gas fees as a marginal sunk cost. Developing fully flash loan-resistant protocols is challenging, as it would deny large-quantity orders with differently sourced capital. We will cover mitigation scenarios in Section VI.

*2) Other Market Attacks:* A rebase mechanism in a clone of Yearn called SoftYearn was exploited to obtain a large amount of funds [43]. This mechanism was implemented to adjust an elastic token supply based on demand. After manipulating the rebase, the attacker was able to sell his tokens for the previous price as Uniswap's token price did not account for the rebase, which also makes this a market attack.

### B. Protocol Attacks

Within our dataset, 6 protocols had vulnerabilities affecting internal protocol security, without involvement of other DeFi protocols. Of these, 2 were exploited by adversaries, the others reported by security researchers. Protocol attacks to individual protocols are generally performed to gain partial or full control over a protocol's governance.

In 2 protocols, a vulnerability existed to manipulate token balances. The September 2020 attack on bZx allowed attackers to create and transfer tokens towards them to artificially increase their token balance [44]. Within Opyn, a vulnerability in its contract allowed double-spending of tokens due to a faulty *loop* function [42]. In both cases the vulnerabilities were exploited by malicious actors.

TABLE I
DECENTRALIZED FINANCE INCIDENTS

| Date | Project | Type | Contract | Attack | Technique | USD Impact | Source |
|---|---|---|---|---|---|---|---|
| 06 May 19 | Maker | Lending | 0x8e2a84d6ade1e7fffee039a35ef5f19f13057152 | Protocol Attack | Vote Manipulation | - | [29] |
| 13 Jul 19 | 0x | Infrastructure | 0x4f833a24e1f95d70f028921e27040ca56e09ab0b | Protocol Attack | Signature Exploit | - | [30] |
| 18 Sep 19 | DDEX | Interface | 0xeb1f1a285fee2ab60d2910f2786e1d036e09eaa8 | Market Attack | Price Oracle Exploit | - | [31] |
| 27 Sep 19 | AirSwap | Trading | 0x5abcfbd462e175993c6c350023f8634d71daa61d | Protocol Attack | Signature Exploit | - | [32] |
| 30 Sep 19 | bZx | Lending | 0x9ae49c0d7f8f9ef4b864e004fe86ac8294e20950 | Market Attack | Price Oracle Exploit | - | [33] |
| 17 Feb 20 | bZx | Lending | 0x4f4e0f2cb72e718fc0433222768c57e823162152 | Market Attack | Price Oracle Exploit | 298,250 | [34] |
| 18 Feb 20 | bZx | Lending | 0x360f85f0b74326cddff33a812b05353bc537747b | Market Attack | Price Oracle Exploit | 633,000 | [35] |
| 18 Feb 20 | Nexus Mutual | Infrastructure | 0x6a313ff2a3e66db968ee3984bff178973e589322 | Protocol Attack | Vote Manipulation | - | [36] |
| 20 Feb 20 | Nexus Mutual | Infrastructure | 0x6a313ff2a3e66db968ee3984bff178973e589322 | Market Attack | Price Oracle Exploit | - | [36] |
| 12 Mar 20 | Maker | Lending | 0xd8a04f5412223f513dc55f839574430f5ec15531 | Economic Attack | Mempool Manipulation | 8,32m | [37] |
| 17 Apr 20 | Uniswap | Trading | 0x4f30e682d0541eac91748bd38a648d759261b8f3 | Vyper Exploit | Reentrancy | 300,000 | [38] |
| 19 Apr 20 | Lendf.Me | Interfaces | 0xa6a6783828ab3e4a9db54302bc01c4ca73f17efb | Solidity Exploit | Reentrancy | 1.2m | [39] |
| 19 Jun 20 | Bancor | Trading | 0x8dfeb86c7c962577ded19ab2050ac78654fea9f7 | Solidity Exploit | Public Method | 134,691 | [40] |
| 28 Jun 20 | Balancer | Trading | 0x81d73c55458f024cdc82bbf27468a2deaa631407 | Market Attack | Price Oracle Exploit | 50,000 | [41] |
| 04 Aug 20 | Opyn | Infrastructure | 0x951d51baefb72319d9fbe941e1615938d89abfe2 | Protocol Attack | Double Spend | 67,910 | [42] |
| 07 Sep 20 | SoftYearn | Interfaces | 0x88093840aad42d2621e1a452bf5d7076ff804d61 | Market Attack | Rebase Exploit | 250,000 | [43] |
| 13 Sep 20 | bZx | Lending | 0x1d496da96caf6b518b133736beca85d5c4f9cbc5 | Protocol Attack | Circulating Supply | - | [44] |
| 25 Oct 20 | Harvest | Interfaces | 0xcc6028a9fa486f52efd2b95b949ac630d287ce0af | Market Attack | Price Oracle Exploit | 21.53m | [45] |
| 12 Nov 20 | Akropolis | Interfaces | 0x2afa3c8bf33e65d5036cd0f1c3599716894b3077 | Market Attack | Price Oracle Exploit | 2m | [46] |
| 14 Nov 20 | Value | Interfaces | 0x49e833337ece7afe375e44f4e3e8481029218e5c | Market Attack | Price Oracle Exploit | 6m | [47] |

The governance of Maker, AirSwap and Nexus Mutual was vulnerable to manipulation of mechanisms overseeing protocol operations. The vectors have been disclosed by security researchers and thus not maliciously exploited, but could however have impacted on-chain assets. The attack on Maker would have allowed malicious actors to remove user votes and lock user funds forever [29]. 0x was vulnerable to fill orders on behalf of other users, due to a weak implementation of a signature algorithm. AirSwap also had problems with the implementation of signature algorithms. A fault in a feature to delegate swapping to another actor could have allowed unsigned swaps [32]. The attack on Nexus Mutual would have allowed to insert malicious proposals into the voting process and whitelist the proposal to make it appear legitimate [36].

### C. Smart Contract Attacks

A total of 3 DeFi protocols was exploited directly due to weak Solidity or Vyper code with techniques, recorded in the SWC Registry [17], which are generally mitigated when adhering to safe coding practices. Uniswap and Lendf.Me were attack in quick succession, both with a reentrancy attack, pulling funds before a malicious transaction is confirmed or denied. It was the ERC-777 token implementation of both platforms that made the reentrancy attacks possible [38], [39] and the exploit was published a full year before it was used [49]. Bancor was attacked through a public *safeTranserFrom* method in its smart contract, which allows users to transfer funds from one address to another [40]. This method should have had private permissions, allowing only the smart contract itself to call it. Given the fact that these vulnerabilities exist with many secure contract templates available is exemplary of the quick genesis of many DeFi protocols: quick and hasty development driven by a fear of missing out.

### D. Economic/Ethereum Attacks

Our sample has one instance of an economic attack, directly leveraging dynamics on the Ethereum blockchain to attack a DeFi protocol. On Black Thursday 2020, when global stock markets crashed, signaling the beginning of the COVID-19 recession, attackers manipulated Ethereum's mempool of transactions waiting to be mined and confirmed. The attacker deliberately congested the mempool with worthless transactions with low gas fees unlikely to finalize quickly. As Ethereum nodes have an economic incentive to mine transactions with high gas rewards, the mempool became clogged. The attackers then took advantage of the delay caused by them by placing zero-bids on Maker's ETH auction and paying nominal gas fees to front-run their malicious transactions [37]. This is the only economic attack with a listed DeFi project that directly exploited Ethereum blockchain dynamics. However the rapid market turn down of March 12, 2020 is a *black swan* event that emphasizes the importance of threat modeling and taking extreme market circumstances into account to discover 'unknown unknown' threats. Maker responded to the attack by extending the duration of an auction to 6 hours [37]. This serves as an excellent example of how to mitigate threats inherited from underlying layers within DeFi development.

## VI. FRAMEWORK OF DEFI SECURITY THREATS

In this section we present our framework of security threats to DeFi. In our analysis we fundamentally observed that the DeFi threat landscape has four underlying root causes:

- *Protocol security:* individual smart contracts with large holdings are a single point of failure to a protocol's security and thus a target for malicious actors.
- *Oracles:* the reliance of many projects on price feeds delivered by oracles makes them target of exploitation.

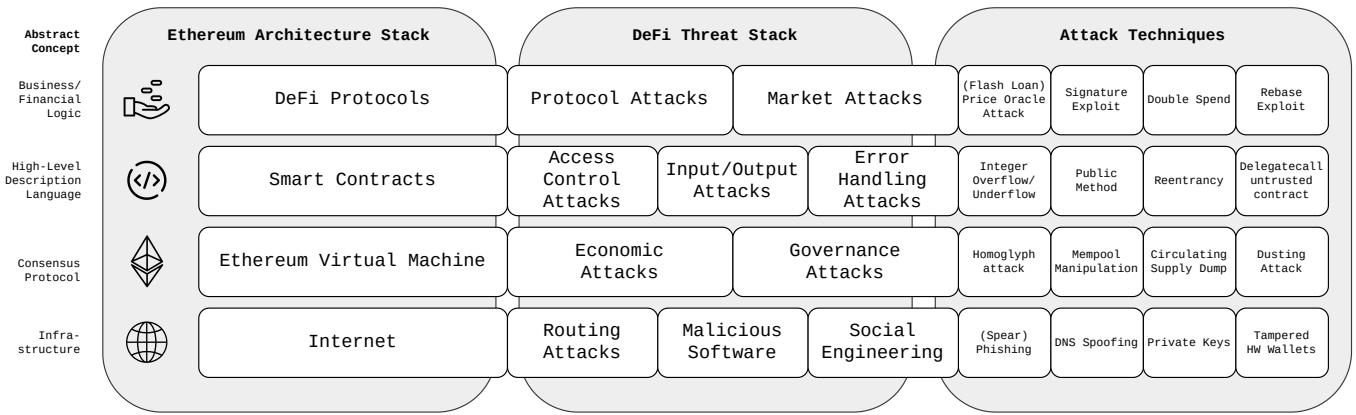

Fig. 2. Framework of Security Threats to DeFi

- *Composability:* while DeFi's 'money lego' architecture is advantageous in many regards, a deficiency in one protocol might cause failure of the whole stack.
- *Lack of custody:* discarding middlemen removes delegation of risk/responsibility prevalent in traditional finance.

These root causes directly affect the security of individual DeFi protocols and the DeFi markets formed by compositions of protocols. However the dynamics of these 'market attacks' and 'protocol attacks' are fundamentally different than than smart contract attacks and the narrative around them can be fuzzy. Our framework structures the narrative around threat mitigation within DeFi. We deem this necessary as the security threats to 'composable' DeFi protocols instead of smart contracts operating in isolation is still a black box to many projects. By standardization of the discussion, a framework promotes better understanding of the threat landscape. An example of this within enterprise security is ATT&CK, which successfully standardized the narrative on adversary TTPs [50].

We present our framework as a stack model because, as pointed out by others, threats on one layer instigate risk to other layers [25]. Figure 2 is a stack representation of how DeFi protocols are supported by smart contracts, which in turn rely on the Ethereum Virtual Machine (EVM), which depends on Internet network, transport and routing protocols. The first column represents the layers of this architecture. Each architecture layer has vulnerabilities of its own, which can be exploited during an attack and thus facilitates threats (second column). Attacks are carried out with a specific technique and take several forms depending on the technique employed. The third column lists techniques observed in incidents in our sample, as well as common examples. Per architecture stack layer, we have identified the following threat categories:

### A. DeFi Protocols

The *business layer* of DeFi is composed by DeFi protocols, vulnerable through the implementation of financial and business logic which serves as their value proposition. DeFi protocols are attacked by *Protocol Attacks* and *Market Attacks*. Protocol Attacks exploit weaknesses in the implementation of a single protocol, such as a protocol's internal governance, with impact is limited to that protocol. Market Attacks exploit compositions of multiple DeFi projects or 'money legos', for example the exploitation of a price oracle having a cascading effect on other projects. While Market Attacks are initiated through exploitation of usually flawed business and financial logic in a single protocol, they potentially have far-reaching effects, impacting multiple protocols. We cover potential mitigation scenarios separately in the next section.

### B. Smart Contracts

Smart contracts are exploited through exploits of high-level languages like Solidity and Vyper for implementing smart contracts, which are *application layer* threats [4], [11], targeting access control, input/output and error handling. Mitigation of integer underflow/overflow, Delegatecall and Floating Pragma threats takes place through safe and secure smart contract programming. Examples are the 'classic' attacks on The DAO and the Parity Wallet, which had weak implementations of Solidity. The application layer threat of contract code exploits is different than the business layer threats of the top DeFi layer. Smart contracts are technically vulnerable to exploitation of software errors and Solidity subtleties, whereas the *business layer* of DeFi is attacked through weakly implemented *business logic*.

To avoid vulnerability to attacks directly exploiting weak smart contract code, projects should implement safe coding practices. Projects should preferably use standard libraries such as SafeMath and community-audited token implementations. Their legitimacy must be verified to avoid tampering by malicious actors. Third-party auditing and publication of reports also has become a best-practice. The number of audits performed and the 'age' of the most recent audit report, as well total engineer weeks spent are useful external metrics. The most recent bZx exploit showed that two audit firms reviewed bZx's code, but failed to find the vulnerability [44]. Besides locating weak Solidity code, audits must focus on business and financial logic. Security audits are never an adequate measure, just like static testing is never a security measure on its own.

## C. Ethereum Virtual Machine

Adhering to Ethereum architecture, EVM is the *consensus layer* [5], [10]. Attacks directly exploit Ethereum blockchain economics and governance on which DeFi projects depend and can potentially impact the Ethereum 'cryptoeconomic' system broader than DeFi (hence *economic* and *governance attacks*). Attacks such as the mempool congestion targeting Maker as discussed in the previous section exploit Ethereum's consensus mechanisms in order to attack DeFi protocols. The execution of 51% attacks at smart contracts overseeing active price oracles is another example of threats on this layer. Fundamentally mitigation of this dependency threat happens in Ethereum core development, but on the level of DeFi protocols developers can account for them in their protocol's business logic. The effects of a 51% attack to alter operational price oracles can partly be mitigated by DeFi projects by implementing a voting system. Furthermore DeFi projects can account for dependency by storing (a part of) their locked value in custody. If the custodian fails to secure funds adequately, losses are offset through insurance reserves.

## D. Internet

The Internet is the *infrastructure* on which Ethereum is built. Routing threats such as DNS spoofing, but also threats such as phishing, key-stealing malware (KryptoCibule, AppleJeus) and social engineering attempts on users acquiring tampered hardware wallets from eBay, put users of DeFi projects and their funds at risk. We include this layer in our framework as these threats have caused significant financial losses token holders in the past [51] and are a threat to the many DeFi projects with centralized admin key storage.

While it is beyond the scope of this paper to provide an overview of such threats and their mitigations, in general a project's degree of trustlessness and decentralization are key towards mitigation. Central private key storage is currently a single point of failure to many projects. Obtained by malicious actors these can be used to unilaterally modify a contract and affect user funds. Time locks and multisignature wallets secure against third-party attacks, but cannot avoid collusion by internal actors. Token-based project governance is promising, but requires controls against majority holdings by the admin team to avoid collusion.

## VII. MARKET ATTACK MITIGATION

This section focuses on the mitigation of flash loan-funded exploits, a market attack vector of growing concern to the DeFi industry. We deconstruct the largest attack to date into a life cycle to promote better understanding of this relatively complex attack vector and its potential mitigation.

### A. The Harvest Attack

Figure 3 shows the attack life cycle of the Harvest attack, with attack phases generic to flash-loan funded price oracle exploits listed in the *Tactic* row. The *Tool* row lists the relevant DeFi service for each phase. Together with *Objective*, this shows how attackers attack compositions of DeFi. The two bottom rows provide a graphical overview of the specific services (mis)used in the Harvest attack, discussed in more detail below, as well the financial impact per phase. It has a *Repeat* phase to account for the additional iterations to circumvent an arbitrage threshold.

Harvest, an automated yield farming service, was attacked on October 26, 2020 [45]. Harvest was exploited through arbitrage assumptions in its financial logic and dependency on price data from a single source, which made it vulnerable to market manipulation. The attacker used funds provided by a flash loan to cause price swings in the underlying Curve liquidity pool. Due to the project's dependence on Curve's Y Pool, the price variation within the pool was reflected in Harvest's share price calculation that takes place during deposit. While Harvest operated a 3% arbitrage check to detect price variations due to large-scale market manipulation, this threshold was too lenient. The attacker simply bypassed it by running multiple cycles causing 1% price swings. The official incident post mortem does not mention how the 3% threshold was established initially [45]. While it is a vector captured logically in threat modeling, according to the project it wasn't recognized in security audits. The attacker seized 24 million USD of the project's funds. For reasons unknown 2.4 million USD were returned to the project, the rest is still missing.

Figure 3 shows how the Harvest attack was executed in the bottom row, but the life cycle as described in the three top rows are generic to all flash loan-funded attacks in our dataset. While the *Repeat* phase was novel to Harvest, it could theoretically take place in any price oracle attack.

### B. Mitigation Scenarios

Below we focus on phases for which Harvest could have denied the attack as part of protocol development.

*1) Swap:* For its pricing Harvest singularly depended on price data derived from Curve, an exchange liquidity pool. This is necessary as Harvest's pool is located in Curve's Y pool and must accurately reflect its price level. Decentralized oracles are not a logical solution as these would also introduce attack surface. However Curve's virtual price feed with price data not derived from a stablecoin was already available and would have mitigated this. In general and for other protocols the time-weighted average price oracle in Uniswap V2 is resistant against manipulation as the upfront cost will exceed return on investment. Maker has a similar mechanism with an Oracle Security Module operating separately from the oracle. A drawback of using oracles resistant against timing attacks, or simply implementing thresholds for extreme price variations, is their slower response to extreme market volatility. For other projects, multiple price feeds can mitigate vulnerability to exploits of single price oracles, a single point of failure to many projects. A decentralized oracle network providing multiple price feeds also provides protection against Sybil attacks, in which an attacker operates multiple oracle nodes to manipulate results.

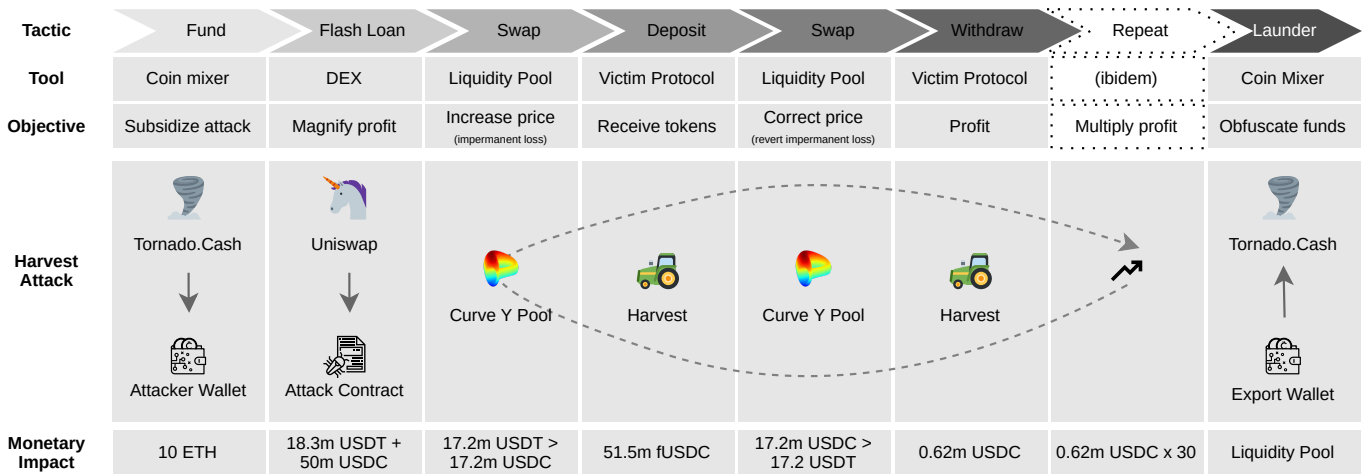

Fig. 3. Life cycle of the loan-funded attack against Harvest, October 26, 2020

*2) Deposit:* with its price driven up, attacker could deposit USDC into Harvest. The 3% arbitrage check was bypassed by making the swaps cause 1% price changes and executing additional cycles to multiply profit with 30. Decreasing the threshold isn't feasible as it would result in false positives for legitimate deposits and withdrawals, however disallowing deposit and withdrawal in a single transaction is a feasible mitigation. Handling the distribution of Harvest shares in a subsequent transaction after withdrawal would also have mitigated the attacker opportunity to cause share fluctuation. In their post-mortem, Harvest suggested a commit-reveal scheme as a potential mitigation. While not truly a commit-reveal scheme as nothing is cryptographically hidden, deposit and distribution of shares would be separated in different blocks, which avoids flash loan-funded attacks [45].

*3) Withdraw:* Harvest operated its own fUSDC tokens which were swappable for stablecoin, which was weaponized by the attacker. Directly depending on the underlying stablecoin would have denied opportunity as the adversary would have to arbitrage the value of his own assets. Like traditional exchanges, DEXes could benefit from traditional approaches such as circuit breakers halting trading during extreme market circumstances. Liquidity and whale alerts to signal suspicious activity can serve as low-cost alerts to inform security monitoring.

This section lists a flash-loan funded price oracle attack. This is a type of market attack, which is mitigated by assuming extreme market behavior in a project's financial logic. Mitigation of a hypothetical attack against a project by front runners colluding with miners to perform malicious transactions over multiple blocks is mitigated on the underlying infrastructure layer. We have presented our framework as a stack to emphasize that inheritance of capabilities from other layers implies inheritance of vulnerabilities - while responsibility can't be shifted. Modeling attack life cycles as part of threat assessments requires developers to take an adversary perspective towards their project and advances understanding of attack vectors and their mitigation.

## VIII. LIMITATIONS

DeFi is a nascent field. While the quantity of incidents in our dataset is relatively limited, it covers significant security incidents with on-chain DeFi. We deemed the threat significant and critical enough to perform this early analysis, considering all incidents occurred over an 18-month timeline. The threat landscape is dynamic and the DeFi field will unequivocally advance in the years ahead, so it remains a work in progress. Our analysis however represents the state-of-the-art in DeFi and can inform security improvement of DeFi protocols.

## IX. CONCLUSION

DeFi is experimental software running in production. Protocols might fail, falling short to generate cash flow and evaporating user funds. The public blockchain on which they rely is a complex and adversarial environment. The freedom it facilitates breeds opportunity for ignorant behavior by irrational and illiterate actors, while each vulnerability will be exploited eventually. With many on-chain protocols still a work in progress, DeFi is a risky cryptoeconomic system.

In this paper we have enumerated security threats to DeFi projects based on in-the-wild attacks, as well as countermeasures to inform mitigation. We have introduced a framework to holistically regard security threats and attacks to DeFi, which we hope contributes to inform better threat modeling and consequent security decision-making in current and future protocols. With institutional interest increasing, security is a key and potentially pivotal responsibility to the ecosystem writ large. The attack vectors covered in this paper need to be addressed, or they will impede DeFi's potential to develop into what it aspires. Similar to *zero trust* in computer network security, DeFi's *'don't trust, verify'* must become default security architecture, rather than a cryptoeconomic ideal.